\documentclass{aastex}

\usepackage{graphics,emulateapj5}

\newbox\grsign \setbox\grsign=\hbox{$>$} \newdimen\grdimen \grdimen=\ht\grsign
\newbox\simlessbox \newbox\simgreatbox
\setbox\simgreatbox=\hbox{\raise.5ex\hbox{$>$}\llap
     {\lower.5ex\hbox{$\sim$}}}\ht1=\grdimen\dp1=0pt
\setbox\simlessbox=\hbox{\raise.5ex\hbox{$<$}\llap
     {\lower.5ex\hbox{$\sim$}}}\ht2=\grdimen\dp2=0pt

\def\simless{\mathrel{\copy\simlessbox}}

\newcommand{\einstein}{\emph{Einstein}}
\newcommand{\rosat}{\emph{ROSAT}}

\newcommand{\mr}[1]{\mathrm{#1}}

\newcommand{\m}{$^{-1}$}

\newcommand{\hhh}{h_{50}}

\newcommand{\lx}{L_\mr{X}}
\newcommand{\lb}{L_\mr{B}}
\newcommand{\lism}{L_\mr{ISM}}
\newcommand{\fism}{f_\mr{ISM}}
\newcommand{\blskip}{}

\begin{document}
\blskip
\submitted{}

\title{The $\lx - \sigma$ Relation for Galaxies and Clusters of Galaxies}

\author{Andisheh Mahdavi\altaffilmark{1} and Margaret J. 
Geller\altaffilmark{2}} \affil{Harvard-Smithsonian
Center for Astrophysics, MS 10, 60 Garden St., Cambridge, MA 02138,
USA}

\altaffiltext{1}{amahdavi@cfa.harvard.edu}
\altaffiltext{2}{mgeller@cfa.harvard.edu}

\shorttitle{$\lx - \sigma$ Relation}
\shortauthors{Mahdavi \& Geller}

\submitted{Submitted April 4, 2001, and accepted May 15, 2001
for publication in \emph{The Astrophysical Journal Letters}}

\begin{abstract}
We demonstrate that individual elliptical galaxies and clusters of
galaxies form a continuous X-ray luminosity---velocity dispersion
($\lx - \sigma$) relation. Our samples of 280 clusters and 57 galaxies
have $\lx \propto \sigma^{4.4}$ and $\lx \propto \sigma^{10}$,
respectively. This unified $\lx - \sigma$ relation spans 8 orders of
magnitude in $\lx$ and is fully consistent with the observed and
theoretical luminosity---temperature scaling laws. Our results support
the notion that galaxies and clusters of galaxies are the luminous
tracers of similar dark matter halos.
\end{abstract}


\section{Introduction}

X-ray observations of galaxy clusters show that gas and dark matter
halos dominate the gravitational potentials of these
$10^{12}$--$10^{16} M_\odot$ systems. Measurements of the temperature
and luminosity of the gas can probe the mass spectrum of clusters and
thus provide a test of cosmological models
\citep{Markevitch98,Henry00}. Crucial to these tests is the idea that
clusters are a homologous set of objects with simple physical
properties: that the least massive clusters are essentially the same
kind of object as the gargantuan, and that simple scaling laws link
observables such the X-ray luminosity $L_X$, temperature $T$, and the
galaxy velocity dispersion $\sigma$ for clusters of all masses.

Basic theoretical models, which assume hydrostatic equilibrium among
the gas, galaxy, and dark matter components, yield $\lx \propto T^2
\propto \sigma^4$ for bremsstrahlung-dominated emission
\citep{Quintana82,Eke98}. Observational tests of these models reveal
several discrepancies. First, for many samples of clusters, data from
both the \einstein\ and \rosat\ satellites yields $\lx \propto
T^{2.5-3} \propto \sigma^{4-5}$ \citep{Mushotzky97,MZ98}. Various
mechanisms---the systematic variation of the cluster baryon fraction
with $T$ \citep{David93}, or cooling flows \citep{Allen98}---may be
responsible for the overall difference in slope.

A further complication, however, is that the deviations from the
simple theoretical models appear to depend on $T$ and $\sigma$, and
thus break the supposed self-similarity of galaxy clusters. A smooth
transition from $\lx \propto T^{2.5}$ to $\lx \propto T^5$ occurs
around $k T \simless 2$ keV \citep{PonmanHCG,Helsdon00}, possibly
because of nongravitational heating of the cluster gas due to, e.g.,
supernova explosions or shocks
\citep*{Kaiser91,Cavaliere98,Tozzi00,Lloyd00}. Oddly, a similar
steepening is not observed in the $\lx - \sigma$ relation
\citep{MZ98,Mahdavi00}.

Resolving the discrepancy between these cluster scaling relations
requires a much larger sample of objects with known $k T \simless 2$
keV and $\sigma \simless 350$ km s\m. The obvious approach is to
undertake an extensive search for poor groups of galaxies; here we
include two recent surveys for X-ray emission from these systems
\citep{Helsdon00,Mahdavi00}. 

Another approach is to regard individual, X-ray emitting elliptical
galaxies as low-mass extensions of clusters. In N-body simulations of
structure formation, dark matter halos are a single class of objects
regardless of mass; galaxies and clusters are just luminous tracers of
the halos. In particular, halos in the simulations of \cite*[NFW]{NFW}
have density profiles with a shape independent of the halo mass. The
NFW profile is consistent with kinematic data for individual
elliptical galaxies \citep{Sato00} as well as with rich cluster data
(\citealt*{Nevalainen00}; \citealt*{Geller99};
\citealt{Rines00}). This agreement suggests that cluster and galaxy
scaling laws are related. As an example, we demonstrate that galaxies
and systems of galaxies form a single, well defined relation in the
$\lx - \sigma$ plane. Throughout the paper we take $H_0 = 50$ km s\m\
Mpc\m.

\begin{figure*}
\begin{center}
\resizebox{6.5in}{!}{\includegraphics{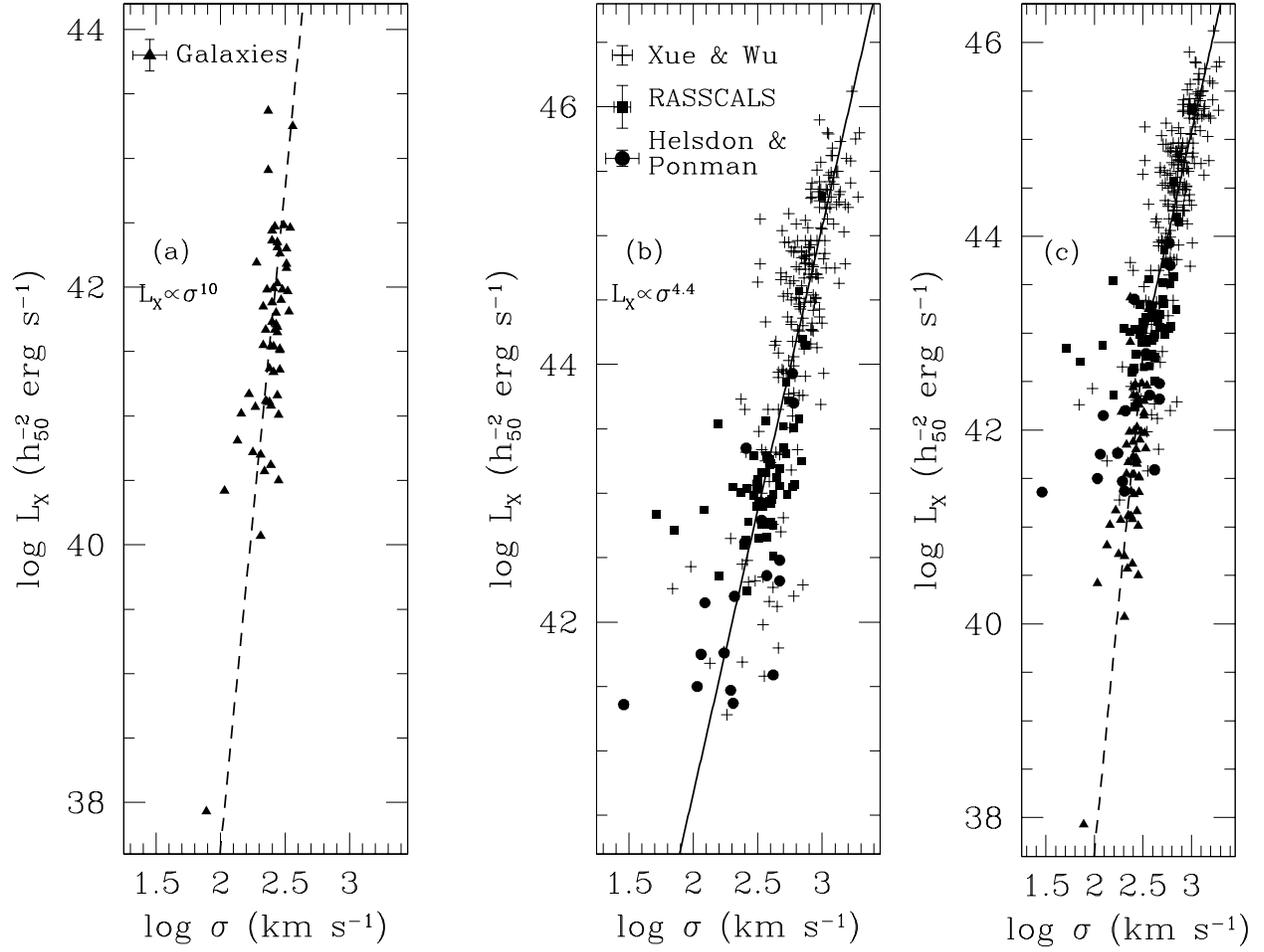}}
\figcaption{$\lx - \sigma$ relation for (a) galaxies and (b)
clusters. The unified relation appears in (c). The dashed and solid
lines show the best fit power laws for the galaxy and cluster samples,
respectively. The error bars show the mean uncertainties in $\lx$ and
$\sigma$ for each sample.
\label{fig:lxsiga}}
\end{center}
\end{figure*}

\section{Sample}

We construct a sample of galaxies and clusters of galaxies that is
suitable for determining the shape of the $\lx - \sigma$ relation over
several decades in luminosity and velocity dispersion. Because the
relation is poorly constrained for the least luminous systems of
galaxies ($\lx < 10^{43}$ erg s\m), we have paid particular attention
to increasing number of poor groups relative to previous analyses.

Our starting point is the catalog of clusters and groups assembled by
\cite*{WuXue99} and \cite{XueWu00}. They gather velocity dispersions
and standardized bolometric X-ray luminosities for 251 systems from
the literature. We use two recent surveys for X-ray emission from
groups \citep{Mahdavi00,Helsdon00} to provide more accurate X-ray
luminosities and to add 29 generally low $\lx$ objects to the sample.
Wherever possible, we also update the velocity dispersions of systems
from our ongoing deep redshift survey of poor groups
\citep{MahdaviGeller99}, which has a limiting magnitude $m_\mr{B} =
16.5$ and which includes typically $\approx 20$ members per system.

The sources for the galaxy sample are \einstein\ and \rosat\ X-ray
observations by \cite*{Eskridge95a} and \cite{Beuing99}.  We match the
galaxies in these catalogs with the line-of-sight velocity dispersions
measured by \cite{Faber89}, obtaining 84 unique objects. We exclude
all galaxies containing active galactic nuclei (AGN) by comparing our
sample with the AGN catalog of \cite{Veron00}.

Another problem is the excess X-ray emission associated with
ellipticals that are the dominant central galaxies in groups
\citep{Helsdon01}. These galaxies have X-ray atmospheres that stand
out from the larger intragroup emission, yet are continuous with
it. In examining their \rosat\ observations of systems containing such
galaxies, \cite{MZ98} and \cite{Helsdon00} argue that the X-ray
emission from dominant ellipticals is linked to the gravitational
potential of the whole group. We therefore exclude all galaxies in
\cite{Eskridge95a} and \cite{Beuing99} with X-ray emission associated
with a cluster or group. The final sample contains 57 uncontaminated
galaxies; the 27 excluded galaxies show no $\lx - \sigma$ correlation.

We have put the galaxy X-ray luminosities on the same bolometric
standard as the clusters, with the assumption that the X-ray emission
is due to a thermal, optically thin plasma with $k T = 1$ keV and a
metallicity equal to $30\%$ the solar value. Because none of the
authors provide error estimates for the galaxy $\lx$ and $\sigma$
measurements, we assign an uncertainty of 30\% to each of these
quantities.

The full catalog, in which we list all 280 clusters and 84 galaxies
with J2000 coordinates, redshifts, and names conforming to the NASA
Extragalactic Database (NED) format, appears in Table \ref{tbl:data},
and is available electronically at http://tdc-www.harvard.edu/lxsigma,
as well as in the arXiv source distribution for this paper.

\section{Discussion}

Here we evaluate the strength and shape of the correlation between
X-ray luminosity and velocity dispersion for both clusters and
individual galaxies. We model the $\lx - \sigma$ relation as a power
law for each sample, and examine its consistency with the observed and
theoretical gas temperature scaling laws.

To conduct a detailed statistical analysis of the two samples, we
first calculate Kendall's $\tau$ as a measure of correlation between
$\lx$ and $\sigma$. We then perform linear regression in log-log space
by minimizing the error-weighted orthogonal distance to the straight
line; see \cite{NR} and \cite{Lupton}.  The results
appear in Table \ref{tbl:stats} and Figure
\ref{fig:lxsiga}. 

Clusters and elliptical galaxies define power laws in the $\lx -
\sigma$ plane. The galaxies have highly correlated $\lx$ and $\sigma$
(with a $10^{-5}$ chance of a false correlation) and a very steep $\lx
\propto \sigma^{10}$. The clusters show a similarly robust
correlation, but a shallower $\lx \propto \sigma^{4.4}$. Table
\ref{tbl:stats} and Figure \ref{fig:lxsigb} show that the galaxies
have a much smaller scatter around the best fit line than do the
clusters.

The galaxy scatter---0.1 dex on the average---is similar to that of
the X-ray fundamental plane constructed by \citet[FP99]{Fukugita99},
who fit a relation of the form $\lx \propto \sigma^a r^b$, where $r$
is the characteristic radius of the elliptical galaxy.  Using a sample
of 30 galaxies, FP99 show that the best fit $a$ and $b$ range from
$0.8-4.0$ and $0.6-2.3$ respectively, depending on the data which
determine $r$ (X-ray or optical). The discrepancy between their $a$
and the $\lx - \sigma$ slopes in Table \ref{tbl:stats} results from
the strong correlation between $r$ and the total system mass.

In contrast to the tightness of the galaxy $\lx - \sigma$ relation,
the behavior of the poorest clusters of galaxies is puzzling. Groups
with $\lx \simless 10^{43}$ erg s\m\ show only a weak correlation in
the $\lx - \sigma$ plane ($\tau = 0.2$, with a 1\% chance of a false
result). The scatter in the cluster $\lx - \sigma$ relation falls from
0.252 dex to 0.182 dex when we exclude poor groups. The groups with
the largest fit residuals tend to be too bright for their velocity
dispersions, an effect previously described by \cite*{ian} and
\cite{Mahdavi00}. 

The large scatter in the $\lx - \sigma$ relation for galaxy groups may
result from a nonequilibrium galaxy velocity distribution or from
unresolved sources within the intragroup gas. \emph{XMM} observations
of the compact group HCG 16 \citep{Turner01}, for example, show that
the bulk of the X-ray emission from this system is associated with the
individual galaxies; a robust estimate of $\lx$ for systems like HCG
16 depends on a careful separation of the diffuse intragroup plasma
from the galaxy emission. Thus deep optical redshift surveys and high
resolution X-ray observations of more galaxy groups are necessary to
reveal the nature of the large scatter in the $\lx - \sigma$ relation.

The unified $\lx - \sigma$ relation (Figure \ref{fig:lxsiga}c) is
physically consistent with the $\lx - T$ and $\sigma - T$ relations
for individual ellipticals and systems of galaxies. According to the
cluster compilation of \cite{XueWu00},
\begin{eqnarray}
\lx \propto T^{2.8\pm0.1} & & \mr{for\ } \lx > 10^{43} \mr{\ erg s}^{-1}; \\
\lx \propto T^{5.6\pm1.8} & & \mr{for\ } \lx < 10^{43} \mr{\ erg s}^{-1}; \\
T \propto \sigma^{1.5\pm0.2} & & \mr{for\ all\ } T
\end{eqnarray}
A compilation of elliptical galaxy temperatures by \cite{Davis96} has
$T \propto \sigma^{1.45\pm0.2}$, consistent with the cluster
result. Eliminating $T$ from the above relations yields $L_X \propto
\sigma^{4.2\pm0.4}$ and $L_X \propto \sigma^{8.4\pm2.9}$ for systems
above and below the $10^{43}$ erg s\m\ break. These predictions match
the results of our analysis (Table \ref{tbl:stats}).

\resizebox{3in}{!}{\includegraphics{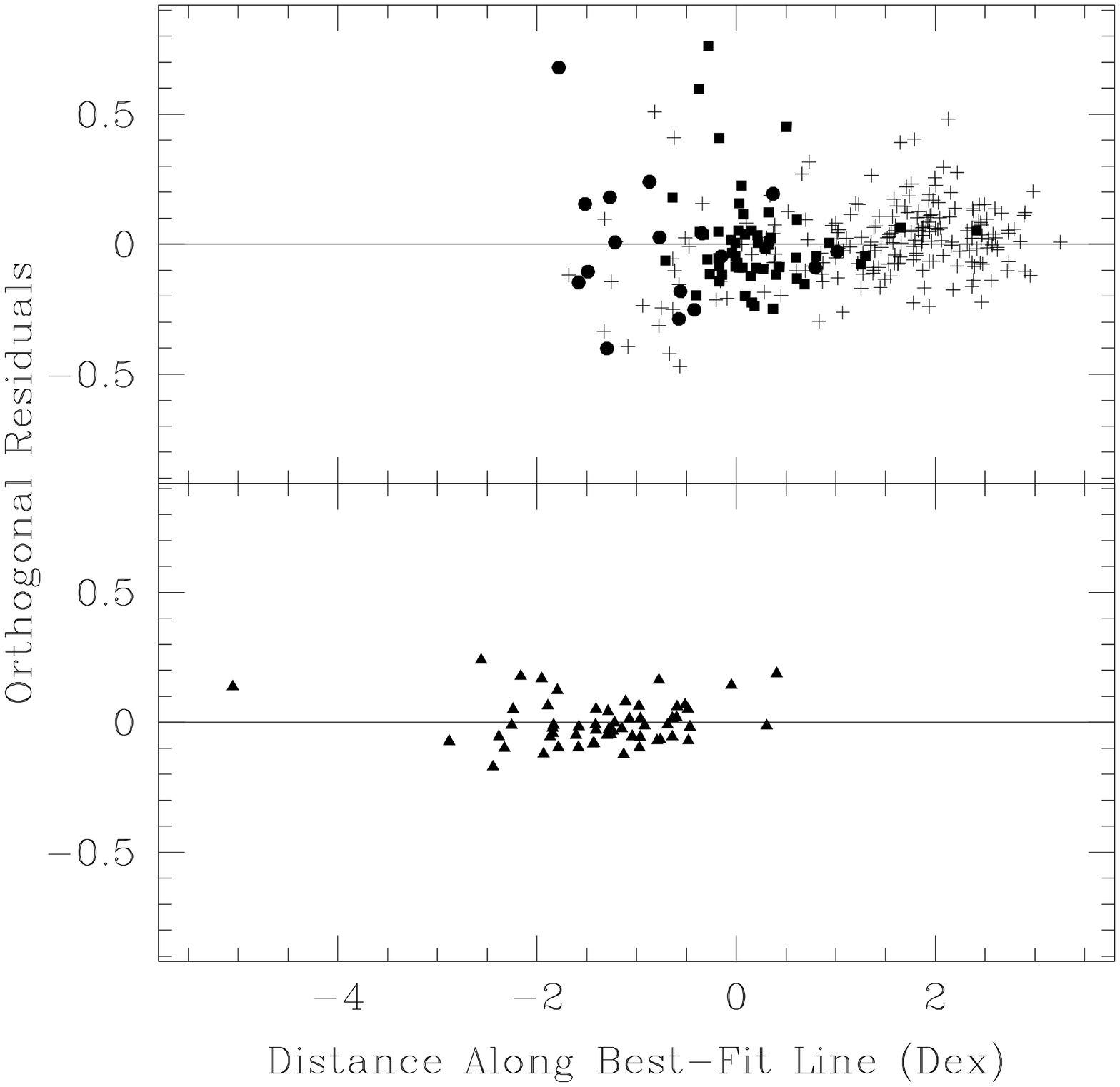}}
\figcaption{(a,b) Orthogonal $\lx - \sigma$ residuals for clusters
(top) and individual galaxies (bottom). The zero point of the x-axis
corresponds to the intersection of the two best-fit power laws at
$\sigma = 330$ km s\m, $\lx = 10^{43}$ erg s\m.
\label{fig:lxsigb}}

\vspace{0.2in}

Differences in the gas content of ellipticals are a concern in
interpreting these relations. The \emph{Einstein} and \emph{ROSAT}
observations of X-ray faint ellipticals---those with the lowest ratios
of X-ray to optical luminosity---are probably dominated by integrated,
unresolved emission from low-mass X-ray binaries (LMXB).  For example,
\emph{Chandra} observations of NGC 4697 resolve 60\% of the emission
into LMXBs, with another 20\% due to unresolved LMXBs and only 20\% of
the emission due to the interstellar gas \citep*{Irwin00}. On the
other hand, the bulk of the emission in the brightest ellipticals is
from the hot ISM. Because both the gas content and the number of LMXBs
of a galaxy should be related to its total mass, it is not surprising
that we observe a tight $\lx - \sigma$ relation despite these physical
differences.

Here we attempt to adjust the $\lx - \sigma$ relation for differences
in the contribution of the galaxy ISM to the total X-ray
luminosity. We assume that the total X-ray emission fraction from the
ISM, $\fism$, (1) is a function of the X-ray to optical emission
ratio, $\lx/\lb$, and (2) is anchored at 20\% by NGC 4697 and at 99\%
by NGC 5044, the brightest uncontaminated galaxy in our sample, where
the emission is almost entirely from the hot ISM \citep{Buote00}.
Using the $\lx/\lb$ data in \cite{Eskridge95a}, these assumptions
yield
\begin{equation}
\fism \propto \left(\frac{\lx}{\lb}\right)^{0.3 \pm 0.1}.
\label{eq:fism}
\end{equation}
\cite{Eskridge95a} observe a significant correlation between the X-ray
luminosity and X-ray to optical emission ratio: $\lx/\lb \propto
\lx^{0.56\pm0.04}$.  Combining this result with equation (\ref{eq:fism}) and
our $\lx - \sigma$ fit, we find
\begin{equation}
\lism = \fism \lx \propto \lx^{1 + 0.56 \times 0.3} \propto \sigma^{12
\pm 5}.
\end{equation}
We conclude that the variation in the gas content of elliptical
galaxies has only a mild effect on the galaxy $\lx - \sigma$ relation,
steepening it within the measured uncertainties.

\section{Conclusion}

Clusters of galaxies and elliptical galaxies form a continuous,
well-defined relation in the $\lx - \sigma$ plane. The best-fit power
laws have the form $\lx \propto \sigma^m$, with $m =
4.4^{+0.7}_{-0.3}$ and $m = 10.2^{+4.1}_{-1.6}$ respectively,
intersecting at $\sigma = 330$ km s\m, $\lx = 10^{43}$ erg s\m. The
steepening of the $\lx - \sigma$ relation from clusters to individual
galaxies supports models where the gaseous medium is preheated by
supernova explosions or merging shocks, and is consistent with the
observed $\lx - T$ and $\sigma - T$ relations for galaxies and systems
of galaxies.  The systematic variation in the gas content of
elliptical galaxies has a negligible effect on these results.

The scatter in the $\lx - \sigma$ relation is smallest at the scale of
objects which are more likely to have reached dynamical
equilibrium---rich clusters of galaxies and individual
ellipticals. Poor groups of galaxies have the largest scatter, an
indication that unresolved, embedded X-ray sources or a
nonequilibrium galaxy velocity distribution affect the integrated
properties of these systems.

We thank the anonymous referee for useful comments. This research has
been supported by the Smithsonian Institution.

\begin{deluxetable}{lrrrrrrcccccl}
\tabletypesize{\small}
\tablecaption{Cluster and Galaxy Catalog \label{tbl:data}}
\tablehead{\colhead{NED} & \multicolumn{3}{c}{RA} &
\multicolumn{3}{c}{DEC} & $c z$ & \multicolumn{2}{c}{$\log{\sigma}$} & 
\multicolumn{2}{c}{$\log{\lx}$\tablenotemark{b}} \\
\colhead{Object\tablenotemark{a}} & \multicolumn{3}{c}{J2000} &
\multicolumn{3}{c}{J2000} & km s\m & \multicolumn{2}{c}{km s\m} & 
\multicolumn{2}{c}{$\hhh^{-2}$ erg s\m} & \colhead{Comments
\tablenotemark{c}}}
\startdata
\cutinhead{Clusters}
      ABELL2717 &00 &02 &59.4 &-36 &02 &06 &0.049 &2.73 & 0.04 &44.37 & 0.07& \\
      ABELL2721 &00 &06 &14.5 &-34 &42 &51 &0.115 &2.91 & 0.04 &44.93 & 0.09& \\
      ABELL2734 &00 &11 &20.1 &-28 &52 &19 &0.062 &2.80 & 0.04 &44.76 & 0.06& \\
        ABELL13 &00 &13 &38.5 &-19 &30 &19 &0.094 &2.95 & 0.04 &44.70 & 0.10& \\
      ABELL2744 &00 &14 &19.5 &-30 &23 &19 &0.308 &3.29 & 0.07 &45.80 & 0.10& \\
        SRGB062 &00 &18 &25.2 &+30 &04 &13 &0.023 &2.59 & 0.05 &43.26 & 0.14& R S34-115 \\
    CL0016+1609 &00 &18 &33.3 &+16 &26 &36 &0.541 &3.09 & 0.05 &45.45 & 0.05& CL0016+16 \\
        ABELL21 &00 &20 &30.8 &+28 &37 &39 &0.095 &2.79 & 0.13 &44.91 & 0.09& \\
        SRGB063 &00 &21 &38.4 &+22 &24 &20 &0.019 &2.53 & 0.05 &43.16 & 0.18& R S49-1479 \\
ZWCL0024.0+1652 &00 &26 &36.0 &+17 &08 &36 &0.390 &3.13 & 0.08 &44.63 & 0.09& CL0024+16 \\
        SRGB075 &00 &41 &15.3 &+25 &28 &54 &0.015 &1.71 & 0.13 &42.84 & 0.22& R \\
        ABELL85 &00 &41 &37.8 &-09 &20 &33 &0.056 &2.91 & 0.04 &45.29 & 0.03& \\
     ABELLS0084 &00 &49 &18.8 &-29 &31 &36 &0.110 &2.52 & 0.06 &44.78 & 0.08& S84 \\
       ABELL115 &00 &55 &59.5 &+26 &19 &14 &0.197 &3.07 & 0.13 &45.49 & 0.10& \\
       ABELL119 &00 &56 &21.4 &-01 &15 &47 &0.044 &2.94 & 0.07 &44.86 & 0.05& \\
      NGC315GRP &00 &58 &25.0 &+30 &39 &11 &0.016 &2.09 & 0.13 &42.15 & 0.15& H \\
      NGC383GRP &01 &00 &41.2 &+32 &11 &23 &0.017 &2.67 & 0.13 &43.31 & 0.13& \\
       ABELL133 &01 &02 &39.0 &-21 &57 &15 &0.057 &2.87 & 0.05 &44.82 & 0.05& \\
        SRGB087 &01 &04 &29.8 &-00 &44 &52 &0.018 &2.42 & 0.08 &42.24 & 0.33& R \\
      NGC383GRP &01 &07 &27.7 &+32 &23 &59 &0.017 &2.72 & 0.03 &43.31 & 0.02& HR S34-111 SRGB090 \\
       ABELL151 &01 &08 &52.3 &-15 &25 &01 &0.053 &2.85 & 0.04 &44.36 & 0.07& \\
       APMCC147 &01 &08 &57.9 &-46 &07 &50 &0.023 &3.01 & 0.05 &43.93 & 0.01& CL0107$-$46 \\
      ABELL2877 &01 &09 &49.3 &-45 &54 &02 &0.025 &2.87 & 0.03 &43.91 & 0.07& \\
       ABELL154 &01 &10 &58.1 &+17 &39 &56 &0.064 &2.93 & 0.11 &44.26 & 0.02& \\
       ABELL160 &01 &12 &51.4 &+15 &30 &54 &0.045 &2.76 & 0.13 &43.94 & 0.13& \\
        SS2B037 &01 &13 &53.2 &-31 &44 &21 &0.018 &2.47 & 0.06 &42.98 & 0.10& R \\
       ABELL168 &01 &15 &09.8 &+00 &14 &51 &0.045 &2.64 & 0.03 &44.18 & 0.07& \\
        S34-113 &01 &23 &19.1 &+33 &21 &39 &0.019 &2.81 & 0.04 &43.34 & 0.13& \\
       ABELL189 &01 &23 &40.4 &+01 &38 &38 &0.033 &2.41 & 0.12 &43.32 & 0.10& \\
      NGC524GRP &01 &24 &01.6 &+09 &27 &37 &0.008 &2.31 & 0.13 &41.37 & 0.11& H \\
       ABELL193 &01 &25 &07.3 &+08 &41 &36 &0.049 &2.86 & 0.04 &44.48 & 0.09& \\
      NGC533GRP &01 &25 &29.1 &+01 &48 &17 &0.018 &2.61 & 0.04 &42.95 & 0.02& RH SRGB102 \\
       ABELL194 &01 &25 &33.1 &-01 &30 &25 &0.018 &2.53 & 0.06 &43.34 & 0.08& \\
        SRGB103 &01 &26 &01.3 &-01 &22 &02 &0.017 &2.60 & 0.04 &43.23 & 0.05& R \\
          HCG12 &01 &27 &33.7 &-04 &40 &14 &0.049 &2.43 & 0.13 &42.31 & 0.08& \\
        SS2S056 &01 &36 &54.4 &-14 &01 &22 &0.038 &2.67 & 0.09 &43.06 & 0.14& R \\
       ABELL222 &01 &37 &27.4 &-12 &58 &45 &0.213 &2.76 & 0.13 &44.88 & 0.02& \\
        SRGB115 &01 &49 &15.1 &+13 &04 &09 &0.016 &2.49 & 0.08 &42.90 & 0.10& R \\
       ABELL262 &01 &52 &50.4 &+36 &08 &46 &0.016 &2.72 & 0.03 &43.93 & 0.04& \\
       ABELL272 &01 &55 &19.1 &+33 &56 &41 &0.088 &2.84 & 0.10 &44.87 & 0.09& \\
        SRGB119 &01 &56 &21.6 &+05 &37 &04 &0.018 &2.57 & 0.05 &42.66 & 0.03& HR S49-140 NGC741GRP\\
          HCG15 &02 &07 &39.0 &+02 &08 &18 &0.023 &2.66 & 0.13 &41.80 & 0.12& \\
          HCG16 &02 &09 &31.3 &-10 &09 &31 &0.013 &2.13 & 0.13 &41.68 & 0.06& \\
        SS2B085 &02 &29 &12.8 &-10 &51 &20 &0.014 &2.20 & 0.11 &42.36 & 0.26& R \\
        SRGB145 &02 &31 &48.0 &+01 &16 &27 &0.022 &2.59 & 0.05 &42.94 & 0.31& R \\
       ABELL370 &02 &39 &50.5 &-01 &35 &08 &0.375 &3.13 & 0.06 &45.32 & 0.04& \\
        SS2B101 &02 &49 &34.1 &-31 &11 &10 &0.021 &2.58 & 0.06 &43.27 & 0.06& R ABELLS301 \\
        SRGB155 &02 &52 &48.7 &-01 &17 &02 &0.023 &2.47 & 0.05 &43.29 & 0.20& R \\
         WBL088 &02 &54 &32.2 &+41 &35 &10 &0.017 &2.94 & 0.05 &44.48 & 0.07& AWM7 \\
           A400 &02 &57 &37.4 &+06 &00 &45 &0.023 &2.74 & 0.05 &43.72 & 0.05& R SRGB161 \\
       ABELL399 &02 &57 &56.4 &+13 &00 &59 &0.072 &2.98 & 0.03 &45.90 & 0.05& \\
       ABELL401 &02 &58 &56.9 &+13 &34 &56 &0.074 &3.06 & 0.03 &45.42 & 0.04& \\
       ABELL407 &03 &01 &43.7 &+35 &49 &48 &0.046 &2.78 & 0.13 &44.06 & 0.11& \\
        SS2B110 &03 &04 &11.1 &-12 &05 &03 &0.012 &2.56 & 0.09 &42.76 & 0.34& R \\
  MS0302.5+1717 &03 &05 &19.0 &+17 &28 &38 &0.425 &2.81 & 0.06 &44.96 & 0.05& MS0302+16 \\
      ABELL3093 &03 &10 &52.0 &-47 &23 &43 &0.083 &2.64 & 0.07 &43.32 & 0.04& \\
      ABELL3112 &03 &17 &52.4 &-44 &14 &35 &0.075 &2.74 & 0.06 &45.17 & 0.04& \\
       ABELL426 &03 &18 &36.4 &+41 &30 &54 &0.018 &3.11 & 0.03 &45.50 & 0.01& \\
        S49-142 &03 &20 &43.7 &-01 &03 &07 &0.021 &1.84 & 0.01 &42.26 & 0.14& \\
      ABELL3126 &03 &28 &43.7 &-55 &42 &44 &0.086 &3.02 & 0.06 &44.86 & 0.07& \\
      ABELL3128 &03 &30 &34.6 &-52 &33 &12 &0.060 &2.92 & 0.02 &44.68 & 0.04& \\
      ABELL3158 &03 &42 &39.6 &-53 &37 &50 &0.060 &2.99 & 0.03 &45.06 & 0.03& \\
       ABELL458 &03 &45 &50.6 &-24 &16 &42 &0.106 &2.87 & 0.04 &44.78 & 0.09& \\
       ABELL400 &03 &55 &33.3 &-36 &34 &18 &0.320 &2.78 & 0.05 &43.78 & 0.09& \\
      ABELL3223 &04 &08 &34.5 &-30 &49 &08 &0.060 &2.81 & 0.04 &44.27 & 0.01& \\
       ABELL478 &04 &13 &20.7 &+10 &28 &35 &0.088 &2.96 & 0.10 &45.51 & 0.06& \\
     NGC1587GRP &04 &30 &46.1 &+00 &24 &25 &0.012 &2.03 & 0.13 &41.50 & 0.18& H \\
      ABELL3266 &04 &31 &11.9 &-61 &24 &23 &0.059 &3.06 & 0.03 &45.22 & 0.02& \\
       ABELL496 &04 &33 &37.1 &-13 &14 &46 &0.033 &2.84 & 0.05 &44.83 & 0.04& \\
  MS0440.5+0204 &04 &43 &09.7 &+02 &10 &19 &0.190 &2.78 & 0.04 &44.87 & 0.06& MS0440+02 \\
       ABELL514 &04 &47 &40.0 &-20 &25 &44 &0.071 &2.95 & 0.04 &44.51 & 0.08& \\
  MS0451.6-0305 &04 &54 &10.9 &-03 &01 &07 &0.550 &3.14 & 0.03 &45.73 & 0.13& MS0451+02 \\
       ABELL520 &04 &54 &19.0 &+02 &56 &49 &0.199 &2.99 & 0.03 &45.57 & 0.10& \\
     ABELLS0506 &05 &01 &04.4 &-24 &24 &41 &0.320 &3.11 & 0.10 &45.24 & 0.03& CL0500$-$24 \\
          HCG33 &05 &10 &47.9 &+18 &02 &05 &0.026 &2.24 & 0.13 &41.77 & 0.11& \\
       ABELL539 &05 &16 &35.1 &+06 &27 &14 &0.028 &2.92 & 0.04 &44.27 & 0.05& \\
      ABELL3360 &05 &40 &18.8 &-43 &23 &31 &0.085 &2.92 & 0.05 &44.32 & 0.04& \\
       ABELL548 &05 &47 &01.7 &-25 &36 &59 &0.042 &2.93 & 0.03 &44.42 & 0.13& \\
      ABELL3376 &06 &00 &43.6 &-40 &03 &00 &0.046 &2.86 & 0.04 &44.67 & 0.04& \\
      ABELL3389 &06 &21 &47.4 &-64 &57 &35 &0.027 &2.77 & 0.04 &43.68 & 0.03& \\
      ABELL3391 &06 &26 &15.4 &-53 &40 &52 &0.051 &2.90 & 0.04 &44.70 & 0.04& \\
      ABELL3395 &06 &27 &31.1 &-54 &23 &58 &0.051 &2.92 & 0.02 &44.77 & 0.03& \\
       ABELL569 &07 &09 &10.4 &+48 &37 &10 &0.020 &2.51 & 0.09 &43.48 & 0.09& \\
       ABELL576 &07 &21 &24.1 &+55 &44 &20 &0.039 &2.98 & 0.04 &44.43 & 0.07& \\
       ABELL578 &07 &25 &01.3 &+66 &59 &07 &0.087 &2.90 & 0.13 &44.78 & 0.03& \\
     NGC2276GRP &07 &43 &05.5 &+85 &22 &30 &0.006 &2.40 & 0.13 &43.65 & 0.13& NGC2300 \\
     NGC2563GRP &08 &20 &24.4 &+21 &05 &46 &0.016 &2.53 & 0.13 &42.79 & 0.02& H \\
       ABELL671 &08 &28 &29.3 &+30 &25 &01 &0.050 &3.00 & 0.13 &44.32 & 0.09& \\
       ABELL665 &08 &30 &45.2 &+65 &52 &55 &0.182 &3.08 & 0.13 &45.62 & 0.07& \\
        NRGB004 &08 &38 &11.5 &+25 &07 &00 &0.029 &2.19 & 0.05 &43.54 & 0.14& R \\
ZWCL0839.9+2937 &08 &42 &56.5 &+29 &27 &59 &0.194 &2.87 & 0.06 &44.96 & 0.06& MS0839+29 \\
          HCG35 &08 &45 &19.5 &+44 &31 &18 &0.054 &2.54 & 0.13 &42.35 & 0.11& \\
       ABELL744 &09 &07 &17.6 &+16 &39 &53 &0.073 &2.91 & 0.07 &44.27 & 0.03& \\
       ABELL754 &09 &08 &50.1 &-09 &38 &12 &0.054 &3.03 & 0.10 &45.36 & 0.03& \\
       ABELL750 &09 &09 &06.7 &+11 &01 &48 &0.180 &3.28 & 0.03 &45.30 & 0.09& \\
          HCG37 &09 &13 &35.6 &+30 &00 &51 &0.022 &2.65 & 0.06 &42.12 & 0.06& \\
        NRGS027 &09 &16 &06.2 &+17 &36 &08 &0.029 &2.59 & 0.06 &43.22 & 0.18& R \\
           A779 &09 &19 &48.0 &+33 &45 &32 &0.023 &2.57 & 0.04 &43.29 & 0.14& R NRGB032 \\
        NRGS038 &09 &23 &35.0 &+22 &19 &58 &0.031 &2.79 & 0.04 &43.07 & 0.24& R \\
        N45-384 &09 &27 &51.8 &+30 &01 &55 &0.027 &2.38 & 0.13 &42.45 & 0.12& \\
        NRGB045 &09 &33 &27.1 &+34 &03 &02 &0.027 &1.85 & 0.11 &42.71 & 0.42& R \\
       ABELL851 &09 &42 &56.6 &+46 &59 &22 &0.407 &3.03 & 0.08 &45.21 & 0.02& \\
          HCG42 &10 &00 &13.1 &-19 &38 &24 &0.013 &2.32 & 0.13 &42.20 & 0.03& H NGC3091 \\
        NRGS076 &10 &06 &41.8 &+14 &25 &49 &0.030 &2.51 & 0.04 &42.94 & 0.30& R \\
ZWCL1006.1+1201 &10 &08 &46.4 &+11 &47 &16 &0.221 &2.96 & 0.05 &44.96 & 0.07& MS1006+12 \\
  MS1008.1-1224 &10 &10 &34.1 &-12 &39 &48 &0.301 &3.02 & 0.04 &44.96 & 0.06& MS1008$-$12 \\
        NRGB078 &10 &13 &53.5 &+38 &44 &24 &0.023 &2.53 & 0.05 &42.76 & 0.52& R N56-393 \\
       ABELL957 &10 &13 &57.3 &-00 &54 &54 &0.044 &2.82 & 0.05 &44.26 & 0.08& \\
       ABELL963 &10 &17 &09.6 &+39 &01 &00 &0.206 &3.04 & 0.14 &45.35 & 0.08& \\
     NGC3258GRP &10 &23 &32.8 &-34 &45 &30 &0.010 &2.60 & 0.13 &43.24 & 0.03& \\
   HYDRACLUSTER &10 &36 &51.3 &-27 &31 &35 &0.013 &2.79 & 0.03 &43.91 & 0.04& A1060 \\
          HCG48 &10 &37 &45.6 &-27 &04 &50 &0.009 &2.55 & 0.13 &41.58 & 0.14& \\
      ABELL1069 &10 &39 &54.3 &-08 &36 &40 &0.065 &2.56 & 0.11 &44.33 & 0.10& \\
        SS2B153 &10 &50 &26.9 &-12 &50 &26 &0.015 &2.31 & 0.13 &43.05 & 0.07& R \\
  MS1054.4-0321 &10 &57 &00.2 &-03 &37 &27 &0.823 &3.07 & 0.06 &45.30 & 0.13& MS1054$-$03 \\
          A1142 &11 &00 &50.9 &+10 &33 &17 &0.035 &2.70 & 0.05 &43.35 & 0.20& R NRGS110 \\
      ABELL1146 &11 &01 &20.6 &-22 &43 &08 &0.142 &3.01 & 0.04 &44.82 & 0.07& \\
          A1185 &11 &10 &31.4 &+28 &43 &39 &0.033 &2.82 & 0.03 &43.58 & 0.13& R NRGS117 \\
      ABELL1213 &11 &16 &29.1 &+29 &15 &37 &0.047 &2.74 & 0.12 &43.77 & 0.13& \\
     NGC3607GRP &11 &17 &55.9 &+18 &07 &35 &0.004 &2.62 & 0.13 &41.59 & 0.03& H \\
          HCG51 &11 &22 &21.6 &+24 &19 &41 &0.028 &2.73 & 0.12 &42.99 & 0.13& R NRGB128 \\
        SS2B164 &11 &22 &43.0 &-07 &44 &54 &0.024 &2.56 & 0.05 &42.93 & 0.27& R \\
     NGC3665GRP &11 &23 &30.6 &+38 &43 &31 &0.007 &1.46 & 0.13 &41.36 & 0.10& H \\
      ABELL1300 &11 &32 &00.7 &-19 &53 &34 &0.307 &3.08 & 0.13 &45.68 & 0.12& \\
      ABELL1291 &11 &32 &04.4 &+56 &01 &26 &0.053 &2.96 & 0.13 &44.13 & 0.02& \\
      ABELL1314 &11 &34 &48.7 &+49 &02 &25 &0.034 &2.82 & 0.09 &43.76 & 0.10& \\
          HCG57 &11 &37 &50.5 &+21 &59 &06 &0.030 &2.54 & 0.04 &41.98 & 0.22& \\
          HCG58 &11 &42 &09.4 &+10 &16 &30 &0.021 &2.41 & 0.13 &42.64 & 0.19& R NRGB151 \\
          A1367 &11 &44 &44.6 &+19 &41 &59 &0.022 &2.87 & 0.03 &44.15 & 0.02& R NRGB155 \\
      ABELL1377 &11 &46 &57.9 &+55 &44 &20 &0.051 &2.69 & 0.13 &43.96 & 0.02& \\
     NGC4065GRP &12 &04 &09.5 &+20 &13 &18 &0.024 &2.62 & 0.04 &42.99 & 0.04& HR N79-299A NRGB177 \\
     NGC4073GRP &12 &04 &21.7 &+01 &50 &19 &0.020 &2.78 & 0.13 &43.70 & 0.01& H \\
        N79-299 &12 &05 &51.2 &+20 &32 &19 &0.023 &2.62 & 0.07 &42.76 & 0.07& \\
        NRGB184 &12 &08 &01.0 &+25 &15 &13 &0.023 &2.54 & 0.09 &42.90 & 0.12& R \\
      ABELL1507 &12 &15 &50.3 &+59 &58 &20 &0.060 &2.37 & 0.13 &43.73 & 0.14& \\
     NGC4261GRP &12 &20 &02.3 &+05 &20 &24 &0.007 &2.67 & 0.13 &42.32 & 0.02& H \\
     NGC4325GRP &12 &23 &18.2 &+10 &37 &19 &0.025 &2.41 & 0.13 &43.35 & 0.03& H \\
   VIRGOCLUSTER &12 &26 &32.1 &+12 &43 &24 &0.004 &2.83 & 0.03 &44.18 & 0.01& VIRGO \\
  MS1224.7+2007 &12 &27 &14.9 &+19 &51 &06 &0.327 &2.90 & 0.05 &44.91 & 0.03& MS1224+20 \\
  MS1231.3+1542 &12 &33 &52.4 &+15 &25 &34 &0.238 &2.82 & 0.04 &44.74 & 0.10& MS1231+15 \\
        N79-284 &12 &35 &58.6 &+26 &55 &29 &0.025 &2.78 & 0.13 &42.20 & 0.16& \\
     NGC4636GRP &12 &42 &57.2 &+02 &31 &34 &0.004 &2.67 & 0.13 &42.48 & 0.01& H \\
CENTAURUSCLUSTER&12 &48 &51.8 &-41 &18 &21 &0.011 &2.77 & 0.03 &44.18 & 0.02& A3526 \\
        SS2B187 &12 &50 &34.1 &-14 &26 &15 &0.014 &2.39 & 0.09 &42.60 & 0.20& R \\
          A1631 &12 &52 &50.6 &-15 &25 &37 &0.015 &2.65 & 0.12 &43.12 & 0.08& R SS2B189 \\
     NGC4761GRP &12 &52 &57.9 &-09 &09 &26 &0.015 &2.56 & 0.05 &43.16 & 0.01& HR SS2B191 HCG62 \\
      ABELL3528 &12 &54 &18.2 &-29 &01 &16 &0.053 &2.99 & 0.04 &43.69 & 0.13& \\
      ABELL1644 &12 &57 &14.8 &-17 &21 &13 &0.047 &2.88 & 0.03 &44.85 & 0.05& \\
      ABELL3532 &12 &57 &19.2 &-30 &22 &13 &0.055 &2.87 & 0.06 &44.74 & 0.06& \\
      ABELL1651 &12 &59 &22.9 &-04 &11 &10 &0.084 &2.98 & 0.06 &45.27 & 0.05& \\
    COMACLUSTER &12 &59 &48.7 &+27 &58 &50 &0.023 &3.00 & 0.02 &45.31 & 0.01& NRGB226 \\
ZWCL1305.4+2941 &13 &07 &50.0 &+29 &25 &44 &0.241 &2.91 & 0.06 &44.16 & 0.13& CL1322+30 \\
      ABELL1689 &13 &11 &34.2 &-01 &21 &56 &0.183 &3.26 & 0.05 &45.75 & 0.07& \\
     NGC5044GRP &13 &14 &22.7 &-16 &32 &04 &0.008 &2.56 & 0.13 &43.25 & 0.13& \\
        NRGS241 &13 &20 &16.6 &+33 &08 &13 &0.037 &2.70 & 0.04 &43.52 & 0.12& R \\
      ABELL3556 &13 &24 &06.2 &-31 &39 &38 &0.048 &2.76 & 0.06 &44.33 & 0.11& \\
     NGC5129GRP &13 &24 &36.0 &+13 &55 &40 &0.023 &2.43 & 0.06 &42.78 & 0.04& HR NRGB244 \\
  *GHO1322+3027 &13 &24 &48.2 &+30 &11 &34 &0.751 &3.20 & 0.07 &45.22 & 0.13& CYGNUS$-$A \\
      ABELL1736 &13 &26 &52.1 &-27 &06 &33 &0.046 &2.72 & 0.09 &44.68 & 0.06& \\
      ABELL3558 &13 &27 &54.8 &-31 &29 &32 &0.048 &2.87 & 0.03 &45.12 & 0.04& \\
     NGC5171GRP &13 &29 &22.3 &+11 &47 &31 &0.023 &2.60 & 0.04 &42.92 & 0.05& HR MKW11NRGB247 \\
     SC1327-312 &13 &29 &47.0 &-31 &36 &29 &0.050 &2.76 & 0.09 &44.50 & 0.13& SC1327-312 \\
      ABELL3559 &13 &29 &53.9 &-29 &31 &29 &0.046 &2.66 & 0.06 &43.65 & 0.03& \\
      ABELL3562 &13 &33 &31.8 &-31 &40 &23 &0.049 &2.87 & 0.03 &44.79 & 0.05& \\
        NRGB251 &13 &34 &25.7 &+34 &40 &54 &0.025 &2.42 & 0.05 &43.04 & 0.17& R \\
      ABELL1767 &13 &36 &00.3 &+59 &12 &43 &0.070 &2.97 & 0.09 &44.67 & 0.05& \\
 RXJ1340.6+4018 &13 &40 &33.4 &+40 &17 &48 &0.171 &2.58 & 0.26 &43.65 & 0.13& \\
      ABELL1775 &13 &41 &55.6 &+26 &21 &53 &0.072 &3.18 & 0.12 &44.78 & 0.06& \\
        SS2S237 &13 &43 &58.8 &-19 &50 &32 &0.033 &2.62 & 0.10 &42.75 & 0.31& R \\
      ABELL3571 &13 &47 &28.9 &-32 &51 &57 &0.039 &3.02 & 0.04 &45.26 & 0.03& \\
      ABELL1795 &13 &49 &00.5 &+26 &35 &07 &0.063 &2.92 & 0.04 &45.41 & 0.03& \\
          HCG67 &13 &49 &03.5 &-07 &12 &20 &0.024 &2.38 & 0.13 &41.69 & 0.10& SS2B239 \\
      ABELL1800 &13 &49 &41.4 &+28 &04 &08 &0.075 &2.86 & 0.13 &44.84 & 0.06& \\
     NGC5353GRP &13 &51 &37.0 &+40 &32 &12 &0.008 &2.24 & 0.13 &41.76 & 0.03& H \\
      ABELL1809 &13 &53 &18.9 &+05 &09 &15 &0.079 &2.88 & 0.04 &44.56 & 0.10& \\
          HCG68 &13 &53 &40.9 &+40 &19 &07 &0.008 &2.26 & 0.13 &41.28 & 0.27& \\
      ABELL1831 &13 &59 &10.2 &+27 &59 &28 &0.061 &2.50 & 0.13 &44.64 & 0.06& \\
ZWCL1358.1+6245 &13 &59 &54.3 &+62 &30 &36 &0.328 &2.97 & 0.03 &45.34 & 0.08& MS1358+62 \\
          3C295 &14 &11 &20.6 &+52 &12 &09 &0.464 &3.22 & 0.09 &45.41 & 0.05& 3C295 \\
      ABELL1904 &14 &22 &07.9 &+48 &33 &22 &0.071 &2.86 & 0.07 &44.12 & 0.02& \\
      ABELL1913 &14 &26 &51.8 &+16 &40 &34 &0.053 &2.66 & 0.10 &43.90 & 0.02& \\
        NRGB302 &14 &28 &33.1 &+11 &22 &07 &0.026 &2.51 & 0.04 &42.65 & 0.59& R N67-309 \\
         WBL514 &14 &34 &00.9 &+03 &46 &52 &0.029 &2.76 & 0.20 &43.18 & 0.12& MKW7 \\
      ABELL1940 &14 &35 &27.6 &+55 &08 &58 &0.140 &2.73 & 0.11 &44.55 & 0.03& \\
         WBL518 &14 &40 &43.1 &+03 &27 &11 &0.027 &2.63 & 0.08 &44.15 & 0.06& MKW8 \\
        N56-381 &14 &47 &06.6 &+11 &35 &29 &0.029 &2.42 & 0.04 &42.48 & 0.13& \\
        NRGS317 &14 &47 &09.8 &+13 &42 &23 &0.030 &2.50 & 0.05 &43.11 & 0.30& R \\
      CL1446+26 &14 &49 &28.2 &+26 &07 &57 &0.370 &3.17 & 0.13 &45.03 & 0.08& CL1447+26 \\
      ABELL1983 &14 &52 &44.0 &+16 &44 &46 &0.044 &2.74 & 0.05 &44.03 & 0.10& \\
      ABELL1991 &14 &54 &30.2 &+18 &37 &51 &0.059 &2.82 & 0.11 &44.47 & 0.07& \\
  MS1455.0+2232 &14 &57 &15.2 &+22 &20 &30 &0.258 &3.05 & 0.05 &45.47 & 0.02& MS1455+22 \\
      ABELL2009 &15 &00 &15.2 &+21 &22 &09 &0.153 &2.91 & 0.13 &45.36 & 0.08& \\
          HCG73 &15 &02 &40.1 &+23 &21 &13 &0.045 &1.98 & 0.13 &42.43 & 0.11& \\
     NGC5846GRP &15 &05 &47.0 &+01 &34 &25 &0.006 &2.57 & 0.13 &42.36 & 0.02& H \\
      ABELL2029 &15 &10 &58.7 &+05 &45 &42 &0.077 &3.07 & 0.03 &45.62 & 0.03& \\
      ABELL2040 &15 &12 &45.2 &+07 &25 &48 &0.046 &2.66 & 0.12 &43.94 & 0.02& \\
  MS1512.4+3647 &15 &14 &25.1 &+36 &36 &30 &0.372 &2.84 & 0.06 &44.88 & 0.10& MS1512+36 \\
      ABELL2052 &15 &16 &45.5 &+07 &00 &01 &0.035 &2.75 & 0.06 &44.63 & 0.03& \\
      ABELL2061 &15 &21 &15.3 &+30 &39 &17 &0.078 &2.74 & 0.08 &44.95 & 0.06& \\
         MKW03S &15 &21 &50.7 &+07 &42 &18 &0.045 &2.78 & 0.04 &44.65 & 0.05& MKW3S \\
      ABELL2065 &15 &22 &42.6 &+27 &43 &21 &0.073 &3.04 & 0.12 &45.13 & 0.05& \\
      ABELL2063 &15 &23 &01.8 &+08 &38 &22 &0.035 &2.82 & 0.03 &44.57 & 0.04& NRGS341 \\
      ABELL2069 &15 &23 &57.9 &+29 &53 &26 &0.116 &2.92 & 0.13 &45.30 & 0.06& \\
      ABELL2079 &15 &28 &04.7 &+28 &52 &40 &0.066 &2.83 & 0.06 &44.43 & 0.08& \\
         WBL574 &15 &32 &29.3 &+04 &40 &54 &0.040 &2.53 & 0.13 &43.00 & 0.13& MKW9 \\
      ABELL2092 &15 &33 &19.4 &+31 &08 &58 &0.067 &2.70 & 0.08 &43.95 & 0.08& \\
      ABELL2107 &15 &39 &47.9 &+21 &46 &21 &0.041 &2.76 & 0.11 &44.31 & 0.07& \\
      ABELL2124 &15 &44 &59.3 &+36 &03 &40 &0.066 &2.91 & 0.04 &44.49 & 0.10& \\
      ABELL2142 &15 &58 &16.1 &+27 &13 &29 &0.091 &3.05 & 0.04 &45.79 & 0.03& \\
      ABELL2147 &16 &02 &17.2 &+15 &53 &43 &0.035 &3.03 & 0.09 &44.75 & 0.03& \\
      ABELL2152 &16 &05 &22.4 &+16 &26 &55 &0.041 &2.85 & 0.04 &43.76 & 0.13& \\
      ABELL2162 &16 &12 &30.0 &+29 &32 &23 &0.032 &2.56 & 0.07 &43.32 & 0.10& \\
      ABELL2163 &16 &15 &34.1 &-06 &07 &26 &0.203 &3.23 & 0.13 &46.12 & 0.06& \\
        NRGS385 &16 &17 &15.4 &+34 &55 &00 &0.031 &2.78 & 0.03 &43.51 & 0.09& R \\
          CID64 &16 &18 &00.0 &+35 &06 &00 &0.030 &2.77 & 0.13 &43.60 & 0.13& ZW1615+35 \\
        NRGS388 &16 &23 &01.0 &+37 &55 &21 &0.033 &2.67 & 0.08 &43.19 & 0.08& R \\
  MS1621.5+2640 &16 &23 &35.7 &+26 &33 &50 &0.426 &2.90 & 0.03 &44.91 & 0.08& MS1621+26 \\
      ABELL2197 &16 &28 &10.4 &+40 &54 &26 &0.031 &2.75 & 0.06 &43.51 & 0.08& \\
          HCG82 &16 &28 &22.1 &+32 &49 &25 &0.036 &2.85 & 0.13 &42.29 & 0.14& \\
      ABELL2199 &16 &28 &37.0 &+39 &31 &28 &0.030 &2.90 & 0.04 &44.85 & 0.02& \\
          HCG83 &16 &35 &40.9 &+06 &16 &12 &0.053 &2.70 & 0.13 &42.81 & 0.12& \\
      ABELL2218 &16 &35 &54.0 &+66 &13 &00 &0.176 &3.14 & 0.06 &45.34 & 0.04& \\
        NRGS404 &16 &58 &02.4 &+27 &51 &42 &0.035 &2.56 & 0.09 &43.56 & 0.05& R \\
      ABELL2244 &17 &02 &44.0 &+34 &02 &48 &0.097 &3.09 & 0.13 &45.40 & 0.04& \\
      ABELL2256 &17 &03 &43.5 &+78 &43 &03 &0.058 &3.13 & 0.02 &45.26 & 0.02& \\
      ABELL2255 &17 &12 &31.0 &+64 &05 &33 &0.081 &3.09 & 0.05 &45.09 & 0.02& \\
     NGC6338GRP &17 &15 &21.4 &+57 &22 &43 &0.028 &2.77 & 0.13 &43.93 & 0.01& H N34-175 \\
 RXJ1716.6+6708 &17 &16 &49.6 &+67 &08 &30 &0.813 &3.18 & 0.05 &45.24 & 0.02& RXJ1716.6+6708 \\
      ABELL2271 &17 &17 &17.5 &+78 &01 &00 &0.058 &2.66 & 0.13 &44.09 & 0.03& \\
 RXJ1736.4+6804 &17 &36 &27.7 &+68 &04 &31 &0.026 &2.46 & 0.13 &43.06 & 0.13& \\
      ABELL2280 &17 &43 &06.6 &+63 &44 &46 &0.326 &2.98 & 0.18 &45.22 & 0.13& \\
 RXJ1755.8+6236 &17 &55 &42.8 &+62 &37 &40 &0.027 &2.59 & 0.13 &43.09 & 0.13& \\
 RXJ1756.5+6512 &17 &56 &31.5 &+65 &12 &57 &0.027 &2.29 & 0.13 &42.65 & 0.13& \\
     ABELLS0805 &18 &47 &14.4 &-63 &19 &45 &0.014 &2.67 & 0.08 &43.30 & 0.13& S805 \\
          HCG85 &18 &50 &22.3 &+73 &21 &00 &0.039 &2.62 & 0.13 &42.27 & 0.11& \\
      ABELL2319 &19 &20 &45.3 &+43 &57 &43 &0.056 &3.19 & 0.02 &45.60 & 0.02& \\
          HCG86 &19 &51 &59.2 &-30 &49 &34 &0.020 &2.48 & 0.13 &42.32 & 0.14& \\
      ABELL3651 &19 &52 &10.9 &-55 &05 &16 &0.060 &2.80 & 0.04 &44.52 & 0.18& \\
      ABELL3667 &20 &12 &30.1 &-56 &49 &00 &0.056 &2.99 & 0.02 &45.36 & 0.08& \\
      ABELL3693 &20 &34 &22.0 &-34 &29 &40 &0.091 &2.68 & 0.07 &44.66 & 0.16& \\
      ABELL3695 &20 &34 &47.7 &-35 &49 &39 &0.089 &2.89 & 0.03 &45.06 & 0.07& \\
      ABELL3716 &20 &51 &16.7 &-52 &41 &43 &0.046 &2.98 & 0.05 &44.37 & 0.08& \\
      ABELL3733 &21 &01 &50.0 &-28 &06 &54 &0.038 &2.78 & 0.06 &43.99 & 0.10& \\
      ABELL3744 &21 &07 &13.8 &-25 &28 &54 &0.038 &2.71 & 0.05 &44.95 & 0.08& \\
        SS2S261 &21 &11 &34.6 &-23 &08 &52 &0.033 &2.84 & 0.09 &43.25 & 0.10& R \\
  MS2137.3-2353 &21 &40 &12.8 &-23 &39 &27 &0.313 &2.98 & 0.13 &45.42 & 0.06& MS2137$-$23 \\
      ABELL3809 &21 &46 &57.5 &-43 &54 &06 &0.062 &2.68 & 0.05 &44.71 & 0.08& \\
      ABELL2390 &21 &53 &34.6 &+17 &40 &11 &0.228 &3.04 & 0.02 &45.80 & 0.10& \\
      ABELL3822 &21 &54 &06.2 &-57 &50 &49 &0.076 &2.91 & 0.04 &44.96 & 0.06& \\
      ABELL3825 &21 &58 &22.4 &-60 &23 &39 &0.076 &2.84 & 0.04 &44.61 & 0.09& \\
      ABELL3827 &22 &01 &49.1 &-59 &57 &15 &0.098 &2.98 & 0.18 &45.36 & 0.05& \\
     ABELLS0987 &22 &02 &23.5 &-22 &30 &15 &0.070 &2.83 & 0.07 &44.45 & 0.21& S987 \\
     NGC7176GRP &22 &02 &31.4 &-32 &04 &58 &0.009 &2.29 & 0.13 &41.47 & 0.11& H HCG90 \\
      ABELL2426 &22 &14 &27.3 &-10 &22 &05 &0.098 &2.52 & 0.07 &45.13 & 0.06& \\
        SRGB009 &22 &14 &48.0 &+13 &50 &17 &0.026 &2.50 & 0.06 &43.04 & 0.26& R \\
      ABELL2440 &22 &23 &52.6 &-01 &35 &47 &0.091 &3.00 & 0.07 &45.05 & 0.10& \\
      ABELL3880 &22 &27 &49.5 &-30 &34 &40 &0.058 &2.93 & 0.08 &44.54 & 0.07& \\
      ABELL3888 &22 &34 &23.0 &-37 &43 &29 &0.151 &3.12 & 0.03 &45.50 & 0.07& \\
          HCG92 &22 &35 &57.5 &+33 &57 &36 &0.021 &2.59 & 0.13 &42.16 & 0.04& \\
      ABELL3921 &22 &49 &38.6 &-64 &23 &15 &0.094 &2.69 & 0.09 &45.04 & 0.06& \\
        SRGB013 &22 &50 &00.7 &+11 &40 &15 &0.026 &2.77 & 0.05 &43.05 & 0.21& R S49-146 \\
        SS2B293 &22 &55 &09.8 &-33 &53 &06 &0.028 &2.08 & 0.12 &42.87 & 0.23& R \\
     ABELLS1077 &22 &58 &52.3 &-34 &46 &55 &0.312 &3.22 & 0.05 &45.58 & 0.13& AC114 \\
      ABELL2556 &23 &13 &03.3 &-21 &37 &40 &0.086 &3.10 & 0.09 &44.75 & 0.18& \\
        SRGS034 &23 &18 &31.2 &+18 &41 &52 &0.039 &2.72 & 0.10 &43.86 & 0.06& R \\
        SRGB031 &23 &20 &11.3 &+08 &12 &05 &0.011 &2.60 & 0.06 &42.78 & 0.08& R \\
     NGC7619GRP &23 &20 &32.1 &+08 &22 &26 &0.011 &2.40 & 0.13 &42.62 & 0.02& H \\
      ABELL2589 &23 &24 &00.5 &+16 &49 &29 &0.041 &2.70 & 0.08 &44.53 & 0.05& \\
      ABELL2593 &23 &24 &31.0 &+14 &38 &29 &0.041 &2.85 & 0.06 &44.30 & 0.06& \\
        SRGB037 &23 &28 &46.6 &+03 &30 &49 &0.017 &2.62 & 0.06 &42.51 & 0.44& R \\
      ABELL4010 &23 &31 &10.3 &-36 &30 &26 &0.096 &2.80 & 0.08 &45.09 & 0.14& \\
      ABELL2626 &23 &36 &31.0 &+21 &09 &36 &0.055 &2.82 & 0.06 &44.51 & 0.06& \\
      ABELL2634 &23 &38 &18.4 &+27 &01 &37 &0.031 &2.85 & 0.05 &44.20 & 0.05& SRGS040 \\
      ABELL2657 &23 &44 &51.0 &+09 &08 &40 &0.040 &2.82 & 0.13 &44.45 & 0.05& \\
          HCG97 &23 &47 &24.0 &-02 &19 &08 &0.023 &2.37 & 0.05 &43.01 & 0.20& R SS2B312 \\
      ABELL4038 &23 &47 &31.1 &-28 &12 &10 &0.030 &2.95 & 0.06 &44.52 & 0.03& SS2B313 \\
      ABELL2666 &23 &50 &56.2 &+27 &08 &41 &0.027 &2.68 & 0.07 &42.70 & 0.09& \\
     NGC7777GRP &23 &53 &33.0 &+28 &34 &42 &0.023 &2.06 & 0.13 &41.75 & 0.20& H \\
      ABELL2670 &23 &54 &10.1 &-10 &24 &18 &0.076 &2.96 & 0.03 &44.70 & 0.08& \\
      ABELL4059 &23 &56 &40.7 &-34 &40 &18 &0.048 &2.93 & 0.11 &44.76 & 0.04& \\
\cutinhead{Galaxies}
 MESSIER032 &00& 42 &41.8 &+40 &51 &52 &0.000 &1.89  &0.13 &37.93  &0.13 &E NGC0221 \\
    NGC0315 &00& 57 &48.9 &+30 &21 &09 &0.016 &2.55  &0.13 &42.26  &0.13 &E AGN \\
     IC1625 &01& 07 &42.8 &-46 &54 &24 &0.022 &2.41  &0.13 &41.99  &0.16 &B \\
    NGC0410 &01& 10 &58.9 &+33 &09 &08 &0.018 &2.51  &0.13 &42.30  &0.05 &B \\
    NGC0499 &01& 23 &11.5 &+33 &27 &38 &0.015 &2.37  &0.13 &42.65  &0.13 &E GR S34-113 \\
    NGC0507 &01& 23 &40.0 &+33 &15 &20 &0.016 &2.56  &0.13 &43.25  &0.13 &E \\
    NGC0533 &01& 25 &31.3 &+01 &45 &33 &0.019 &2.50  &0.13 &42.84  &0.13 &E GR NGC533GRP \\
    NGC0708 &01& 52 &46.4 &+36 &09 &07 &0.016 &2.38  &0.13 &43.44  &0.02 &B GR A262\\
    NGC0720 &01& 53 &00.4 &-13 &44 &18 &0.006 &2.39  &0.13 &41.54  &0.13 &E \\
    NGC0741 &01& 56 &21.0 &+05 &37 &44 &0.019 &2.45  &0.13 &42.12  &0.07 &B GR SRGB119\\
    NGC0777 &02& 00 &14.9 &+31 &25 &46 &0.017 &2.54  &0.13 &42.46  &0.05 &B \\
    NGC1052 &02& 41 &04.8 &-08 &15 &21 &0.005 &2.31  &0.13 &40.91  &0.13 &E AGN \\
     IC0310 &03& 16 &42.9 &+41 &19 &30 &0.019 &2.37  &0.13 &42.91  &0.03 &B \\
    NGC1399 &03& 38 &29.3 &-35 &27 &01 &0.005 &2.49  &0.13 &42.48  &0.13 &E \\
    NGC1395 &03& 38 &29.6 &-23 &01 &40 &0.006 &2.41  &0.13 &41.34  &0.13 &E \\
    NGC1404 &03& 38 &52.0 &-35 &35 &34 &0.006 &2.35  &0.13 &41.67  &0.13 &E \\
    NGC1407 &03& 40 &11.8 &-18 &34 &48 &0.006 &2.46  &0.13 &41.51  &0.13 &E \\
    NGC1600 &04& 31 &39.9 &-05 &05 &10 &0.016 &2.51  &0.13 &42.15  &0.13 &E \\
    NGC1573 &04& 35 &04.1 &+73 &15 &45 &0.014 &2.44  &0.13 &41.65  &0.11 &B \\
    NGC2305 &06& 48 &37.3 &-64 &16 &24 &0.012 &2.33  &0.13 &41.85  &0.03 &B \\
    NGC2325 &07& 02 &40.5 &-28 &41 &50 &0.007 &2.13  &0.13 &40.81  &0.14 &B \\
    NGC2329 &07& 09 &08.0 &+48 &36 &56 &0.019 &2.43  &0.13 &42.43  &0.06 &B GR A569\\
    NGC2340 &07& 11 &10.8 &+50 &10 &28 &0.020 &2.40  &0.13 &42.36  &0.08 &B \\
    NGC2300 &07& 32 &19.5 &+85 &42 &33 &0.006 &2.43  &0.13 &41.71  &0.13 &E \\
    NGC2434 &07& 34 &51.4 &-69 &17 &01 &0.005 &2.31  &0.13 &40.07  &0.14 &B \\
    NGC2563 &08& 20 &35.7 &+21 &04 &04 &0.015 &2.42  &0.13 &42.32  &0.13 &E GR NGC2563GRP \\
    NGC2663 &08& 45 &08.0 &-33 &47 &44 &0.007 &2.45  &0.13 &40.50  &0.20 &B \\
    NGC2832 &09& 19 &46.8 &+33 &44 &59 &0.023 &2.56  &0.13 &42.94  &0.13 &E GR A779 \\
    NGC2974 &09& 42 &32.9 &-03 &41 &58 &0.007 &2.35  &0.13 &41.12  &0.13 &E \\
    NGC2986 &09& 44 &16.0 &-21 &16 &41 &0.008 &2.45  &0.13 &41.01  &0.10 &B \\
    NGC3078 &09& 58 &24.6 &-26 &55 &35 &0.008 &2.38  &0.13 &41.37  &0.13 &E \\
    NGC3091 &10& 00 &14.3 &-19 &38 &13 &0.013 &2.46  &0.13 &41.44  &0.09 &B GR HCG42\\
    NGC3226 &10& 23 &27.0 &+19 &53 &54 &0.004 &2.31  &0.13 &40.62  &0.09 &B AGN\\
    NGC3258 &10& 28 &54.1 &-35 &36 &22 &0.009 &2.44  &0.13 &41.69  &0.13 &E \\
    NGC3585 &11& 13 &16.8 &-26 &45 &21 &0.005 &2.34  &0.13 &40.57  &0.13 &E \\
    NGC3607 &11& 16 &54.7 &+18 &03 &06 &0.003 &2.39  &0.13 &41.08  &0.13 &E \\
    NGC3608 &11& 16 &59.0 &+18 &08 &55 &0.004 &2.31  &0.13 &40.70  &0.13 &E \\
    NGC3862 &11& 45 &05.0 &+19 &36 &23 &0.022 &2.42  &0.13 &42.13  &0.15 &B AGN GR A1367\\
    NGC3923 &11& 51 &02.1 &-28 &48 &23 &0.006 &2.33  &0.13 &41.55  &0.13 &E \\
    NGC4073 &12& 04 &27.0 &+01 &53 &48 &0.020 &2.44  &0.13 &43.16  &0.03 &B GR NGC4073GRP\\
    NGC4105 &12& 06 &40.6 &-29 &45 &42 &0.006 &2.38  &0.13 &41.11  &0.13 &E \\
    NGC4125 &12& 08 &05.6 &+65 &10 &29 &0.005 &2.36  &0.13 &41.11  &0.08 &B \\
    NGC4168 &12& 12 &17.3 &+13 &12 &18 &0.008 &2.26  &0.13 &40.50  &0.13 &E AGN \\
    NGC4261 &12& 19 &23.2 &+05 &49 &31 &0.007 &2.47  &0.13 &41.90  &0.13 &E \\
    NGC4278 &12& 20 &06.8 &+29 &16 &51 &0.002 &2.42  &0.13 &39.91  &0.11 &B AGN\\
    NGC4291 &12& 20 &17.7 &+75 &22 &15 &0.006 &2.41  &0.13 &41.54  &0.13 &E \\
    NGC4365 &12& 24 &28.2 &+07 &19 &03 &0.004 &2.39  &0.13 &40.62  &0.13 &E \\
 MESSIER084 &12& 25 &03.7 &+12 &53 &13 &0.004 &2.46  &0.13 &41.36  &0.13 &E NGC4374\\
 MESSIER086 &12& 26 &11.7 &+12 &56 &46&-0.001& 2.40  &0.13 &42.44 & 0.13 &E NGC4406\\
    NGC4458 &12& 28 &57.5 &+13 &14 &31 &0.002 &2.03  &0.13 &40.42  &0.13 &E \\
 MESSIER049 &12& 29 &46.8 &+08 &00 &02 &0.003 &2.46  &0.13 &42.26  &0.13 &E NGC4472\\
    NGC4473 &12& 29 &48.9 &+13 &25 &46 &0.007 &2.25  &0.13 &40.72  &0.13 &E \\
 MESSIER087 &12& 30 &49.4 &+12 &23 &28 &0.004 &2.21  &0.13 &43.08  &0.01 &B AGN NGC4486\\
 MESSIER089 &12& 35 &39.8 &+12 &33 &23 &0.001 &2.42  &0.13 &41.12  &0.13 &E AGN NGC4552\\
    NGC4636 &12& 42 &50.0 &+02 &41 &17 &0.003 &2.28  &0.13 &42.19  &0.13 &E \\
 MESSIER060 &12& 43 &39.6 &+11 &33 &09 &0.004 &2.53  &0.13 &41.81  &0.13 &E NGC4649\\
    NGC4697 &12& 48 &35.7 &-05 &48 &03 &0.004 &2.22  &0.13 &41.17  &0.13 &E \\
    NGC4696 &12& 48 &49.3 &-41 &18 &40 &0.010 &2.35  &0.13 &43.38  &0.01 &B GR CENTAURUS\\
    NGC4760 &12& 53 &07.2 &-10 &29 &39 &0.016 &2.40  &0.13 &41.73  &0.10 &B \\
    NGC4782 &12& 54 &35.8 &-12 &34 &09 &0.013 &2.52  &0.13 &41.97  &0.13 &E \\
    NGC4936 &13& 04 &17.0 &-30 &31 &31 &0.010 &2.40  &0.13 &41.88  &0.10 &B \\
    NGC5044 &13& 15 &24.0 &-16 &23 &06 &0.009 &2.37  &0.13 &43.37  &0.13 &E \\
    NGC5077 &13& 19 &31.6 &-12 &39 &23 &0.009 &2.44  &0.13 &41.27  &0.13 &E AGN \\
    NGC5090 &13& 21 &12.8 &-43 &42 &16 &0.011 &2.43  &0.13 &41.80  &0.11 &B \\
    NGC5129 &13& 24 &10.0 &+13 &58 &36 &0.023 &2.44  &0.13 &42.35  &0.10 &B \\
    NGC5216 &13& 32 &06.9 &+62 &42 &02 &0.010 &2.16  &0.13 &41.02  &0.20 &B \\
     IC4296 &13& 36 &39.4 &-33 &58 &00 &0.012 &2.51  &0.13 &42.18  &0.13 &E \\
    NGC5328 &13& 52 &53.6 &-28 &29 &16 &0.016 &2.44  &0.13 &42.03  &0.07 &B \\
    NGC5419 &14& 03 &38.3 &-33 &58 &50 &0.014 &2.48  &0.13 &41.98  &0.09 &B \\
    NGC5846 &15& 06 &29.2 &+01 &36 &21 &0.006 &2.44  &0.13 &42.31  &0.13 &E \\
    NGC5982 &15& 38 &39.9 &+59 &21 &21 &0.010 &2.42  &0.13 &41.67  &0.13 &E \\
    NGC6160 &16& 27 &41.2 &+40 &55 &36 &0.032 &2.37  &0.13 &42.63  &0.06 &B GR A2197\\
    NGC6166 &16& 28 &38.5 &+39 &33 &06 &0.030 &2.51  &0.13 &44.30  &0.02 &B GR A2199\\
    NGC6173 &16& 29 &44.9 &+40 &48 &42 &0.029 &2.42  &0.13 &42.47  &0.15 &B \\
    NGC6868 &20& 09 &54.1 &-48 &22 &47 &0.010 &2.46  &0.13 &41.52  &0.19 &B \\
    NGC6876 &20& 18 &20.2 &-70 &51 &28 &0.013 &2.36  &0.13 &41.98  &0.13 &E \\
    NGC7196 &22& 05 &55.1 &-50 &07 &11 &0.010 &2.44  &0.13 &41.16  &0.12 &B \\
    NGC7192 &22& 06 &50.3 &-64 &18 &56 &0.010 &2.27  &0.13 &41.07  &0.14 &B \\
    NGC7385 &22& 49 &54.7 &+11 &36 &30 &0.026 &2.41  &0.13 &42.03  &0.21 &B GR SRGB013\\
     IC1459 &22& 57 &10.6 &-36 &27 &44 &0.006 &2.49  &0.13 &41.21  &0.13 &E AGN \\
    NGC7619 &23& 20 &14.4 &+08 &12 &22 &0.013 &2.53  &0.13 &42.20  &0.13 &E GR SRGB031 \\
    NGC7626 &23& 20 &42.3 &+08 &13 &02 &0.011 &2.37  &0.13 &41.70  &0.13 &E GR NGC7619GRP \\
     IC5358 &23& 47 &45.0 &-28 &08 &27 &0.029 &2.29  &0.13 &42.03  &0.07 &B GR A4038\\
    NGC7768 &23& 50 &58.6 &+27 &08 &50 &0.027 &2.38  &0.13 &42.11  &0.14 &B GR A2666\\
\enddata
\tablecomments{All data are from \cite{XueWu00} and \cite{WuXue99} unless
the comments specify otherwise.}
\tablenotetext{a}{Alternate names are indicated in the comment field.}
\tablenotetext{b}{The X-ray luminosities have been standardized to bolometric.}
\tablenotetext{c}{H: X-ray luminosity from \cite{Helsdon00}; R: X-ray luminosity and velocity dispersion from \cite{Mahdavi00}; HR: X-ray luminosity from \cite{Helsdon00}, velocity dispersion from \cite{Mahdavi00}; B: data from \cite{Beuing99}; E: data from \cite{Eskridge95a}; GR: galaxy is coincident with intracluster
emission from the noted system; AGN: galaxy is associated with an active galactic nucleus.}
\end{deluxetable}

\newcommand{\tn}[1]{\tablenotemark{#1}}
\newcommand{\tnt}[1]{\tablenotetext{#1}}
\begin{deluxetable}{lrccccc}
\tablecaption{Statistical Analysis of the $\lx - \sigma$ relation
 \label{tbl:stats}}
\tablehead{
\colhead{Sample}   &
\colhead{$N$} &
\colhead{$\tau$} & \colhead{$P$\tn{a}} &
\colhead{Slope} & 
\colhead{Intercept} & 
\colhead{Scatter\tn{b}}}
\startdata
Clusters     &280 & 0.652 &
$<10^{-6}$&$4.4^{+0.7}_{-0.3}$&$31.8^{+0.9}_{-2.0}$ & 0.252 \\
Galaxies   & 84 & 0.283  & $<10^{-6}$ & 15.3$^{+6.7}_{-2.9}$ &
5.1$^{+7}_{-16}$  & 0.152 \\
$\bullet$ Contaminated  & 27 & 0.163  &0.234  & \nodata & \nodata \\
$\bullet$ Uncontaminated & 57 & 0.377
&$10^{-5}$&10.2$^{+4.1}_{-1.6}$&17.2$^{+3.8}_{-10.1}$ & 0.094  \\
\enddata
\tnt{a}{Probability that an uncorrelated sample would give 
the quoted value of Kendall's $\tau$ or higher by chance.}
\tnt{b}{Root-mean-square orthogonal scatter in the log-log plane.}
\end{deluxetable}

\end{document}